\documentclass[10pt, twocolumn]{article}
                                                                                
\usepackage{amsmath, amssymb}
\usepackage{latexsym}
\usepackage{latex8}
\usepackage{times}
\usepackage{epsfig}
                                                                                
\newenvironment{myitemize}
{\begin{list}{$\bullet$}%
    {\setlength{\itemsep}{-3pt}}%
    {\setlength{\topsep}{0pt}}%
    {\setlength{\partopsep}{0pt}}%
    {\setlength{\parsep}{0pt}}%
}
{\end{list}%
}

\newtheorem{theorem}{Theorem}
\newtheorem{lemma}[theorem]{Lemma}
\newtheorem{proposition}[theorem]{Proposition}
\newtheorem{corollary}[theorem]{Corollary}

\newtheorem{definition}[theorem]{Definition}

\newcommand{\hide}[1]{}
\newcommand{\Reals}{\mbox{$\mathbb R$}}
\newcommand{\embedded}{\prec}

\newcommand{\SA}{\mathsf{SA}}

\renewcommand{\L}{\mathsf{L}} 
\newcommand{\depth}{\mathsf{d}}  
\newcommand{\D}{\mathsf{D}}  
\newcommand{\N}{\mathsf{N}}
\newcommand{\spi}{\mathsf{sumPI}}
\newcommand{\mpi}{\mathsf{maxPI}}
\newcommand{\KI}{\mathsf{KI}}
\newcommand{\PA}{\mathsf{PA}}

\newcommand{\MAJ}[1]{\mathsf{R{-}MAJ}_3^{#1}}
\newcommand{\A}{\mathrm{f_A}}    
\newcommand{\Col}{\mathrm{f_C}}  
\newcommand{\Isum}{\sum_{\genfrac{}{}{0pt}{}{i}{x_i\not=y_i}}}
\newcommand{\rk}{\mathrm{rk}}
\newcommand{\R}{\mathcal{R}}
\newcommand{\SC}{\mathcal{S}}
\newcommand{\ignore}[1]{}

\newenvironment{proof}
{\noindent {\bf Proof:}}
{\nopagebreak{\hfill $\Box$}\\
 \smallskip }
\newenvironment{proofnobox}
{\noindent {\bf Proof:}}
{}
                                                                                
\begin{document}

\title{The quantum adversary method \\and classical formula size lower bounds}
\author{Sophie Laplante \\ LRI, Universit\'e Paris-Sud \\
{\tt laplante@lri.fr} \\
\and Troy Lee \\ CWI and University of Amsterdam \\ {\tt Troy.Lee@cwi.nl} \\
\and Mario Szegedy \\ Rutgers University \\ {\tt szegedy@cs.rutgers.edu}\\
}

\maketitle
\thispagestyle{empty}
\begin{abstract}
We introduce two new complexity measures for Boolean functions, 
which we name $\spi$ and $\mpi$.  The quantity $\spi$ has 
been emerging through a line of research on quantum query complexity lower 
bounds via the so-called quantum adversary 
method~\cite{A02,A03,BSS03,Zha03,LM04}, culminating in
\cite{SS04} with the realization that these many different formulations are 
in fact equivalent.  Given that $\spi$ turns out to be such a robust 
invariant of a function, we begin to investigate this quantity in its own
right and see that it also has applications to classical complexity theory.

As a surprising application we show that $\spi^2(f)$ is a lower bound on the 
formula size, and even, up to a constant multiplicative factor, the 
probabilistic formula size of $f$.  We show that several formula size lower 
bounds in the literature, specifically Khrapchenko and its 
extensions~\cite{K71,Kou93}, including a key lemma of~\cite{H98}, are in fact 
special cases of our method.  The second quantity we introduce, $\mpi(f)$, 
is always at least as large as $\spi(f)$, and is derived from $\spi$ in such 
a way that $\mpi^2(f)$ remains a lower bound on formula size. 

Our main result is proven via a combinatorial lemma which relates the square 
of the spectral norm of a matrix to the squares of the spectral norms of its 
submatrices.  The generality of this lemma gives that our methods 
can also be used to lower bound the communication complexity of relations, 
and a related combinatorial quantity, the rectangle partition number.

To exhibit the strengths and weaknesses of our methods, we look at the
$\spi$ and $\mpi$ complexity of a few examples, including the recursive 
majority of three function, a function defined by Ambainis \cite{A03}, 
and the collision problem. 
\end{abstract}
                                                                                
\section{Introduction}
A central and longstanding open problem in complexity theory is to
prove superlinear lower bounds for the circuit size of an explicit Boolean
function.  While this seems quite difficult, a modest amount of success
has been achieved in the slightly weaker model of formula size, a
formula being simply a circuit where every gate has fan-out at most one.  The
current best formula size lower bound for an explicit function is $n^{3-o(1)}$
by H{\aa}stad~\cite{H98}.
                                                                                
In this paper we show that part of the rich theory developed around
proving lower bounds on quantum query complexity, namely the so-called
quantum adversary argument, can be brought to bear on formula size lower
bounds.  This adds to the growing list of examples of how 
studying quantum computing has led to new results in classical
complexity, including~\cite{SV01,KdW03,A04,LM04}, to cite just a few.

The roots of the quantum adversary argument can be traced to the
hybrid argument of~\cite{BBBV97}, who use it to show a $\Omega(\sqrt{n})$
lower bound on quantum search.  Ambainis developed a more sophisticated 
adversary argument \cite{A02} and later improved this method to the 
full-strength quantum adversary argument \cite{A03}.  Further generalizations 
include Barnum, Saks, and Szegedy~\cite{BSS03} with their
spectral method and Zhang~\cite{Zha03} with his strong adversary method.
Laplante and Magniez~\cite{LM04} use Kolmogorov complexity to capture
the adversary argument in terms of a minimization problem.  
This line of research culminates in recent work of
\v{S}palek and Szegedy~\cite{SS04} who show that in fact all the methods of
\cite{A03,BSS03, Zha03, LM04} are equivalent.
                                                                                
The fact that the quantum adversary argument has so many
equivalent definitions indicates that it is a natural combinatorial property
of Boolean functions which is worthwhile to investigate on its own.  We give
this quantity its own name, $\spi$, and adopt the following primal
formulation of the method, from~\cite{SS04,LM04}.  Letting 
$S \subseteq \{0,1\}^n$ and $f: S \rightarrow \{0,1\}$, be a Boolean function
we say
\begin{equation}
\label{eq:spi}
\spi(f)=\min_{p} \max_{\genfrac{}{}{0pt}{}{x,y}{f(x) \not = f(y)}}
\frac{1}{\Isum \sqrt{p_x(i) p_y(i)}},
\end{equation}
where $p=\{p_x : x\in S\}$ is a family of probability distributions on the
indices $[n]$.  If $Q_{\epsilon}(f)$ is the two sided error quantum query
complexity of $f$ then
$Q_{\epsilon}(f) = \Omega(\spi(f))$.
We show further that $\spi^2(f)$ is a lower bound on the formula 
size of $f$.  Moreover, $\spi^2(f)$ generalizes several formula size lower 
bounds in the literature, specifically Khrapchenko and its 
extensions~\cite{K71,Kou93}, and a key lemma of~\cite{H98} used on the way 
to proving the current best formula size lower bounds for an explicit function. 

We also introduce $$\KI(f) =
    \min_{\alpha\in \Sigma^*}
    \max_{\genfrac{}{}{0pt}{}{x,y}{f(x) \not = f(y)}}
    \min_{i : x_i \neq y_i}
K(i|x,\alpha)+K(i|y,\alpha),$$ where $K$ is prefix-free
Kolmogorov complexity.  This formulation arises from the
quantum and randomized lower bounds of~\cite{LM04}.
This formulation is especially interesting because of the
intuition that it provides.
For example, it allows for a very simple proof that circuit 
depth $\depth(f) \geq \KI(f)$, using
the Karchmer-Wigderson characterization of circuit depth~\cite{KW}.

We define a quantity closely related to $2^{\KI}$, which
we call $\mpi$. 
\begin{equation}
\label{eq:mpi}
\mpi(f)=\min_{p} \max_{\genfrac{}{}{0pt}{}{x,y}{f(x) \not = f(y)}}
\frac{1}{\max_{i:x_i \not = y_i} \sqrt{p_x(i) p_y(i)}}.
\end{equation}
Notice that this is like $\spi$ but where the sum is
replaced by a maximum.  By definition, $\mpi$ is larger than $\spi$, but its 
square is still a lower bound on formula size.
However, $\mpi$ is no longer a lower bound on quantum
query complexity in general, and we give an example of a
partial function $f$ for which $\spi(f)=2$ and $\mpi(f)=\sqrt{n/2}$.
For this function, the collision problem,
$\mpi(f) \gg Q_{\epsilon}(f) = \Theta(n^{1/3})$ \cite{AY04,BHT97}.
                                                                               
We look at several concrete problems to illustrate the strengths
and weaknesses of our methods.
We study the height $h$ recursive majority of three problem, $\MAJ{h}$, and
show that $Q_{\epsilon}(\MAJ{h})=\Omega(2^h)$ and a lower bound of $4^h$
for the formula size.
We also look at a function defined by Ambainis~\cite{A03} to separate
the quantum query complexity of a function from the bound given by the
polynomial method~\cite{BBCMW01}.  This function gives an example where 
$\spi^2$ can give something much better than Khraphchenko's bound.
We also give bounds for the collision problem.

\subsection{Organization}
                                                                                
In Section~\ref{sect:preliminaries}, we give the definitions,
   results, and notation that we use throughout the paper, and
   introduce the quantities $\spi$, $\mpi$, and $\KI$.
In Section~\ref{sect:spi+mpi} we prove some properties of $\spi$ and $\mpi$.
In Section~\ref{sect:formulasize}, we show how 
    $\spi$ and $\mpi$ give rise to formula size lower bounds, 
   for deterministic and probabilistic formula size.  
In Section~\ref{sect:comparison}, we compare our new methods
   with previous methods in formula size complexity.
In Section~\ref{sect:limits}, we investigate the limits of 
   our and other formula lower bound methods.  
Finally, in Section~\ref{sect:applications} we apply our 
    techniques to some concrete problems.
                                                                                
\section{Preliminaries}
\label{sect:preliminaries}

We use standard notation such as $[n]=\{1, \ldots, n\}$, $|S|$ for the 
cardinality of a set $S$, and all logarithms are base 2.   Hamming distance 
is written $d_H$.

\subsection{Complexity measures of Boolean functions}
We use standard measures of Boolean functions, such as 
sensitivity and certificate complexity.  We briefly recall these here, see
\cite{BW02} for more details.  For a set $S \subseteq \{0,1\}^n$ 
and Boolean function $f: S \rightarrow \{0,1\}$, the sensitivity of $f$ on 
input $x$ is the number of positions $i \in [n]$ such that changing the value 
of $x$ in position $i$ changes the function value.  
The zero-sensitivity, written $s_0(f)$ is the maximum over $x \in f^{-1}(0)$ 
of the sensitivity of $f$ on $x$.  The one-sensitivity, $s_1(f)$ is defined
analogously.  The maximum of $s_0(f),s_1(f)$ is the sensitivity of $f$, written
$s(f)$. 

A certificate for $f$ on input $x \in S$ is a subset 
$I \subseteq [n]$ such that for any $y$ satisfying $y_i=x_i$ for all $i \in I$
it must be the case that $f(y)=f(x)$.  The zero-certificate complexity of $f$, 
written $C_0(f)$ is the maximum over all $x \in f^{-1}(0)$ of the minimum 
size certificate of $x$.  Similarly, the one-certificate complexity of $f$, 
written $C_1(f)$ is the maximum over all $x \in f^{-1}(1)$ of the minimum size 
certificate of $x$.

\subsection{Linear algebra}
For a matrix $A$ (respectively, vector $v$) we write $A^T$ (resp. $v^T$) for 
the transpose of $A$, and $A^*$ (resp. $v^*)$ for the 
conjugate transpose of $A$. 
For two matrices $A,B$ we let $A \circ B$ be the Hadamard product of $A$ and 
$B$, that is $(A \circ B)[x,y]=A[x,y]B[x,y]$.  We write $A\ge B$ if $A$ is 
entrywise greater than $B$, and $A \succeq B$ when $A -B$ is positive 
semidefinite, that is $\forall v : v^T(A-B)v \ge 0$.  We let $\rk(A)$ denote
the rank of the matrix $A$.

We will make extensive use of the spectral norm,
denoted $\|A\|_2$.   For a matrix $A$, 
$$
\|A\|_2 = \{ \sqrt{\lambda} : \lambda \mbox{ is the largest eigenvalue of }
A^* A\}.
$$
For a vector $v$, we let $|v|$ be the $\ell_2$ norm of $v$.

We will also make use of the maximum absolute column sum norm, written 
$\|A\|_1$ and defined as $\|A\|_1=\max_j \sum_i | A[i,j] |$, and the 
maximum absolute row sum norm, written $\|A\|_{\infty}$ and defined
$\|A\|_{\infty}= \max_i \sum_j | A[i,j] |$.  

We collect a few facts about the spectral norm.  These can be found in 
\cite{HJ99}.
\begin{proposition}
Let $A$ be an arbitrary $m$ by $n$ matrix.  Then
\begin{enumerate}
\addtolength{\itemsep}{-0.5\baselineskip} 
  \item{$\|A\|_2 = \max_{u,v} \frac{|u^*Av|}{|u| |v|}$}
  \item{$\|A\|_2^2 \le \|A\|_1 \|A\|_{\infty}$}
  \item{For nonnegative matrices $A,B$, if $A \le B$ then 
	$\|A\|_2 \le \|B\|_2$}
\end{enumerate}
\label{pr:fact}
\end{proposition}

\subsection{Deterministic and probabilistic formulae}
A Boolean formula over the standard basis $\{\vee, \wedge, \neg\}$ is a binary
tree where each internal node is labeled with $\vee$ or $\wedge$, and each
leaf is labeled with a literal, that is, a Boolean
variable or its negation.  The size of a formula is its number of leaves.
We naturally identify a formula with the function it computes.
\begin{definition}
\label{df:formula_size}
Let $f: \{0,1\}^n \rightarrow \{0,1\}$ be a Boolean function.  The formula size
of $f$, denoted $\L(f)$, is the size of the smallest formula which computes $f$.
The formula depth of $f$, denoted $\depth(f)$ is the minimum depth of a formula
computing~$f$.
\end{definition}
It is clear that $\L(f) \le 2^{\depth(f)}$; that in fact the opposite inequality
$\depth(f)\le O(\log \L(f))$ also holds is a nontrivial result due to
Spira~\cite{Spi71}.
                                                                                
We will also consider probabilistic formulae, that is, a probability
distribution over deterministic formulae.  We take a worst-case notion
of the size of a probabilistic formula.  Probabilistic formula size
has been studied before, for example in~\cite{Val84,B89,DW97,K04}.
\begin{definition}
\label{df:prob_formula_size}
Let $\{f_j\}_{j \in J}$ be a set of functions with $f_j:S \rightarrow \{0,1\}$
for each $j \in J$.  For a function $f:S\rightarrow \{0,1\}$, we say that $f$ is
$\epsilon$-approximated by $\{f_j\}_{j \in J}$ if there is a probability
distribution $\alpha=\{\alpha_j\}_{j \in J}$ over $J$ such that for every
$x \in S$,
$$
\Pr_{\alpha}[f(x)=f_j(x)] \ge 1- \epsilon.
$$
In particular, if $\max_j \L(f_j) \le s$, then we say that $f$ is
$\epsilon$-approximated by formulas of size $s$, denoted
$\L^{\epsilon}(f) \le s$.
\end{definition}
                                                                                
Note that even if a function depends on all its variables, it is possible that
the probabilistic formula size is less than the number of variables.
                                                                                
\subsection{Communication complexity of relations}
\label{sect:depth}
Karchmer and Wigderson \cite{KW} give an elegant characterization 
of formula
 size in terms of a communication game.  We will use this 
framework to 
present our lower bounds.  This presentation has the 
advantage of 
showing that our methods work more generally for the 
communication complexity
of relations beyond the ``special case'' 
of formula size.  The framework of 
communication complexity also 
allows us to work with the rectangle partition 
number, $C^D(R)$, 
which is known to lower bound communication complexity 
and arises 
very naturally when using our techniques.

Let $X,Y,Z$ be finite sets, and $R \subseteq X {{\times}} Y {{\times}} Z$.  In the
communication game for $R$, Alice is given some $x \in X$, Bob is given some
$y \in Y$ and their goal is to find some $z \in Z$ such that $(x,y,z) \in R$,
if such a $z$ exists.  A communication protocol is
a binary tree where each internal node $v$ is labelled by a either a function 
$a_v : X \rightarrow \{0,1\}$ or $b_v: Y \rightarrow \{0,1\}$ describing 
either Alice's or Bob's message at that node, and where each leaf is labelled
with an element $z \in Z$.  A communication protocol computes $R$ if for all
$(x,y) \in X {\times} Y$ walking down the tree according to $a_v, b_v$ leads
to a leaf labelled with $z$ such that $(x,y,z) \in R$, provided such a $z$ 
exists.  The communication cost $\D(R)$ of $R$ is the height of the smallest 
communication protocol computing $R$.  
The communication matrix of $R$ is the matrix $M_R[x,y] = \{z  : R(x,y,z)\}$.
A rectangle $X' {\times} Y'$ with $X'\subseteq X$ and $Y'\subseteq Y$ 
is monochromatic if $\bigcap_{x\in X', y\in Y'}M_R[x,y] \neq \emptyset$.
The protocol partition number  $C^P(R)$ is the number of leaves in the smallest
communication protocol computing $R$, and the rectangle partition number
$C^D(R)$ is the smallest number of disjoint monochromatic rectangles 
required to cover $X{\times} Y$.  (Note that $C^D(R) \leq C^P(R)$.)

\begin{definition}
For any Boolean function $f$ we associate a relation
$R_f = \{(x,y,i) : f(x)=0, f(y)=1,  x_i \neq y_i  \}$.
\end{definition}

\begin{theorem}[Karchmer-Wigderson]
For any Boolean function $f$, $\L(f)=C^P(R_f)$ and $\depth(f)=\D(R_f)$.
\end{theorem}

\hide{
\begin{proof}[sketch]
We sketch the direction $C^P(R_f) \le \L(f)$, as this will be needed 
later.  Let $\phi$ be a formula for $f$ of size $\L(f)$, and assume without
loss of generality that $\phi$ only has negations at its leaves.  We let
$\phi$ define our communication protocol.  Say that Alice has $x \in f^{-1}(0)$
and Bob has $y \in f^{-1}(1)$.  Initially $\phi(x)=0$ and $\phi(y)$.  We move 
down the tree choosing subformulas which continue to satisfy these conditions. 
Thus for example if node $v$ is labelled by an AND gate, say the AND of 
subformulas $\phi_{v0}, \phi_{v1}$, then
Alice will speak according to the rule $a_v(x) =0$ if $\phi_{v0}(x)=0$ and 
$a(x)=1$ otherwise.  Continuing in this manner with Alice 
speaking at the AND gates and Bob speaking at the $OR$ gates, we will eventually
arrive at a leaf labeled by a single literal $\ell_i$ such that 
$\ell_i(x) \ne \ell_i(y)$.  Thus it must be the case that $x_i \ne y_i$, and
so we label this leaf with index $i$.  It is clear that the number of leaves
of the communication protocol so defined is simply $\L(f)$, and the 
communication cost is the depth of the formula. 
\end{proof}
}
\subsection{$\spi$ and the quantum adversary method}

Knowledge of quantum computing is not needed for reading 
this paper; for completeness, however, we briefly sketch the quantum query 
model.  More background on quantum query complexity and quantum computing
in general can be found in~\cite{BW02, NC00}.

As with the classical counterpart, in the quantum query model we wish to compute
some function $f : S \rightarrow \{0,1\}$, where $S \subseteq \Sigma^n$, and 
we access the input through queries.  The complexity 
of $f$ is the number of queries needed to compute $f$. 
Unlike the classical case, however, we can now make queries in superposition.
Formally, a query $O$ corresponds to the unitary transformation
$$
O: |i,b,z\rangle \mapsto |i,b \oplus x_i, z \rangle
$$
where $i \in [n], b \in \{0,1\}$, and $z$ represents the workspace.
A $t$-query quantum algorithm $A$ has the 
form $A=U_t O U_{t-1}O \cdots O U_1 O U_0$,
where the $U_k$ are fixed unitary transformations independent of the input $x$. 
The computation begins in the state $|0\rangle$, and the result of the 
computation $A$ is the observation of the rightmost bit of $A|0\rangle$. 
We say that $A$ $\epsilon$-approximates~$f$ if the observation of the 
rightmost bit of $A|O\rangle$ is equal to $f(x)$ with probability at least
$1-\epsilon$, for every $x$.  We denote by $Q_\epsilon(f)$ the minimum 
query complexity of a quantum query algorithm which 
$\epsilon$-approximates~$f$.

Along with the polynomial method~\cite{BBCMW01}, one of the main techniques
for showing lower bounds in quantum query complexity is the quantum adversary 
method~\cite{A02,A03,BSS03,Zha03,LM04}.  Recently, \v{S}palek and Szegedy 
\cite{SS04} have shown that all the strong versions of the quantum adversary
method are equivalent, and further that these methods can be nicely 
characterized as primal and dual.

We give the primal characterization as our principal definition of $\spi$.

\begin{definition}[$\spi$]
Let $S \subseteq \{0,1\}^n$ and $f : S \rightarrow \{0,1\}$ be a Boolean 
function.  For every $x \in S$ let $p_x: [n] \rightarrow \mathbb{R}$
be a probability distribution,  that is, $p_x(i) \ge 0$ and
$\sum_i p_x(i)=1$.  Let $p = \{p_x : x \in S\}$.
We define the sum probability of indices to be
$$
\spi(f)=\min_{p} \max_{\genfrac{}{}{0pt}{}{x,y}{f(x) \not = f(y)}}
\frac{1}{\Isum \sqrt{p_x(i) p_y(i)}}
$$
\end{definition}

We will also use two versions of the dual method, both a weight scheme and the
spectral formulation.  The most convenient weight scheme
for us is the ``probability scheme'', given in Lemma~4 of~\cite{LM04}. 

\begin{definition}[Probability Scheme] Let $S \subseteq \{0,1\}^n$ and
$f: S \rightarrow \{0,1\}$ be a Boolean function, and $X=f^{-1}(0), 
Y=f^{-1}(1)$.  Let $q$ be a probability distribution
on $X{\times} Y$, and $p_A,p_B$ be probability distributions on $X,Y$ 
respectively.  Finally let $\{p_{x,i}': x \in X, i \in [n]\}$ and 
$\{p_{y,i}': y \in Y, i \in [n]\}$ be families of probability distributions
on $X,Y$ respectively.  Assume that $q(x,y) = 0$ when $f(x)=f(y)$.  Let 
$P$ range over all possible tuples $(q,p_A, p_B, \{p_{x,i}'\}_{x,i})$ of 
distributions as above. Then
$$
\PA(f)= \max_{P} \min_{\genfrac{}{}{0pt}{}{x,y,i}{f(x)\not=f(y), x_i \not = y_i}} 
\frac{\sqrt{p_A(x)p_B(y)p_{x,i}'(y)p_{y,i}'(x)}}{q(x,y)}
$$
\label{df:prob_scheme}
\end{definition}

We will also use the spectral adversary method.
\begin{definition}[Spectral Adversary] Let $S \subseteq \{0,1\}^n$ and 
$f: S \rightarrow \{0,1\}$ be a Boolean function.  Let 
$X=f^{-1}(0), Y=f^{-1}(1)$.  Let $\Gamma \ne 0$ be 
an arbitrary $|X| {\times} |Y|$ non-negative symmetric matrix that satisfies 
$\Gamma[x,y]=0$ whenever $f(x)=f(y)$.  For $i \in [n]$, let $\Gamma_i$ be the 
matrix:
$$
\Gamma_i[x,y]= \left\{
\begin{array}{ll}
0 & \mbox{ if } x_i=y_i \\
\Gamma[x,y] & \mbox{ if } x_i \not = y_i
\end{array}
\right.
$$
Then
$$
\SA(f)=\max_\Gamma \frac{\|\Gamma\|_2}{\max_i \|\Gamma_i\|_2}
$$
\end{definition}
Note that the spectral adversary method was initially defined \cite{BSS03} for 
symmetric matrices over $X \cup Y$.  The above definition is equivalent:
if $A$ is a $X \cup Y$ matrix satisfying the constraint that $A[x,y]=0$
when $f(x)=f(y)$ then $A$ is of the form 
$A= \left[ \begin{array}{cc}
0 & B \\
B^T & 0 \\ 
\end{array} \right]
$, for some matrix $B$ over $X {\times} Y$.  Then the spectral norm of $A$ is
equal to that of $B$.  Similarly, for any $X {\times} Y$ matrix $A$ we can
form a symmetrized version  of $A$ as above preserving the spectral norm.

We will often use the following theorem implicitly in taking the method
most convenient for the particular bound we wish to demonstrate.
\begin{theorem}[\v{S}palek-Szegedy]
\label{th:SS04}
Let $n \ge 1$ be an integer, $S \subseteq \{0,1\}^n$ and 
$f:S \rightarrow \{0,1\}$.  Then 
$$
\spi(f)=\SA(f)=\PA(f)
$$
\end{theorem}
                                                                                

\subsection{The $\KI$ and $\mpi$ complexity measures}
\label{sect:KI}


The definition of $\KI$ arises from the Kolmogorov complexity
adversary method~\cite{LM04}.
The Kolmogorov complexity $C_U(x)$ of a string $x$,
with respect to a universal Turing machine $U$ is
the length of the shortest program $p$ such that $U(p)=x$.
The complexity of $x$ given $y$, denoted $C(x|y)$ is the
length of the shortest program $p$ such that $U(\langle p, y\rangle)=x$.
When $U$ is such that the set of outputs is prefix-free
(no string in the set is prefix of another in the set),
we write $K_U(x|y)$.  From this point onwards, we fix $U$
and simply write $K(x|y)$. For more background on Kolmogorov 
complexity consult~\cite{LV97}.

\begin{definition}
Let $S \subseteq \Sigma^n$ for an alphabet $\Sigma$.
For any function $f:S \rightarrow \{0,1\}$, let 
$$\KI(f)=    \min_{\alpha\in \Sigma^*}
    \max_{\genfrac{}{}{0pt}{}{x,y}{f(x) \not = f(y)}} 
    \min_{i : x_i \neq y_i}
K(i|x,\alpha)+K(i|y,\alpha).$$
\end{definition}
The advantage of using concepts based on Kolmogorov complexity is that
they often naturally capture the 
information theoretic content of lower bounds.
As an example of this, we give a simple proof  
that $\KI$ is a lower bound on circuit depth.

\begin{theorem}
For any Boolean function $f$,
$\KI(f) \leq \depth(f)$.
\end{theorem}

\begin{proof}
Let $P$ be a protocol for $R_f$.  Fix $x,y$ with different
values under $f$, and let
$T_A$ be a transcript of the messages sent from A to B,
on input $x,y$.
Similarly, let $T_B$ be a transcript of the messages sent from B to A.
Let $i$ be the output of the protocol, with $x_i \neq y_i$.
To print $i$ given $x$, simulate $P$ using $x$ and $T_B$.
To print $i$ given $y$, simulate $P$ using $y$ and $T_A$.
This shows that
$\forall x,y : f(x) \neq f(y), \exists i : x_i \neq y_i,  K(i|x,\alpha) + K(i|y,\alpha) \leq |T_A|+|T_B| \leq \D(R_f)$,
where $\alpha$ is a description of $A$'s and $B$'s algorithms.
\end{proof}

\noindent{\bf Remark}
A similar proof in fact shows that
$ \KI(f)\leq 2 \N(R_f)$, where $N$ is the nondeterministic communication
complexity.  Since the bound does not take advantage of
interaction between the two players, in many cases
we cannot hope to get optimal lower bounds using these techniques.
\medskip

An argument similar to that in~\cite{SS04} shows that 
$$2^{\KI(f)} = \Theta\left(
\min_{p} \max_{\genfrac{}{}{0pt}{}{x,y}{f(x) \not = f(y)}}
\frac{1}{\max_{i} \sqrt{p_x(i) p_y(i)}}\right)
$$
Notice that the right hand side of the equation is identical to 
the definition of $\spi$, except that the sum in the 
denominator is replaced by a maximum.  This 
led us to define the complexity measure $\mpi$, in order
to get stronger formula size lower bounds.

\begin{definition}[$\mpi$]
Let $f:S\rightarrow \{0,1\}$ be a function with $S \subseteq \Sigma^n$. 
For every $x \in S$ let $p_x: [n] \rightarrow \mathbb{R}$
be a probability distribution.  
Let $p = \{p_x : x \in S\}$.
We define the maximum probability of indices to be
$$
\mpi(f)=\min_{p} \max_{\genfrac{}{}{0pt}{}{x,y}{f(x) \not = f(y)}}
\frac{1}{\max_i \sqrt{p_x(i) p_y(i)}}
$$
\end{definition}
It can be easily seen from the definitions that $\spi(f)\le \mpi(f)$ for
any $f$.  The following lemma is also straightforward from the definitions:
\begin{lemma}
\label{le:restriction}
If $S' \subseteq S$ and $f':S' \rightarrow \{0,1\}$ is a domain restriction of
$f:S \rightarrow \{0,1\}$ to $S'$, then $\spi(f') \le \spi(f)$ and
$\mpi(f') \le \mpi(f)$.
\end{lemma}

\section{Properties of $\spi$ and $\mpi$}
\label{sect:spi+mpi}

\subsection{Properties of $\spi$}
Although in general, as we shall see, $\spi$ gives weaker formula size lower 
bounds than $\mpi$, the measure $\spi$ has several nice properties which 
make it more convenient to use in practice.

The next lemma shows that $\spi$ 
behaves like most other complexity measures with
respect to composition of functions:

\begin{lemma}\label{le:upper_composition}
Let $g_{1},\ldots, g_{n}$ be Boolean functions, 
and  $h$ be a function,
$h:\{0,1\}^n\rightarrow \{0,1\}$. If $\spi(g_{j}) \le a$ for $1\le j \le n$ and 
$\spi(h)\le b$, then for
$f = h(g_{1},\ldots,g_{n})$, $\spi(f) \le ab$.
\end{lemma}

\begin{proof}
Let $p$ be an optimal family of
distribution functions associated with $h$ and $p_{j}$
be optimal families of distribution functions associated with $g_{j}$.
Define the distribution function
$$
q_x(i) = \sum_{j\in [n]} p_{g(x)}(j)p_{j,x}(i).
$$
                                                                                
Assume that for $x,y\in S$ we have $f(x)\neq f(y)$.
It is enough to show that
                                                                                
\begin{equation*}
\sum_{i:\; x_{i}\neq y_{i}} 
\sqrt{\sum_{j\in [n]}p_{g(x)}(j)p_{j,x}(i)}
\sqrt{\sum_{j\in [n]}p_{g(y)}(j)p_{j,y}(i)}\\  
\end{equation*}
\begin{equation}
\label{eqc}
\ge \frac{1}{ab}.
\end{equation}
                                                                                
By Cauchy--Schwarz, the left hand side of Eq.~\ref{eqc} is greater than 
or equal to
\begin{equation*}
{\sum_{i:x_{i}\neq y_{i}} \sum_{j\in [n]}
\sqrt{p_{g(x)}(j)p_{j,x}(i)}\sqrt{p_{g(y)}(j)p_{j,y}(i)}}
\end{equation*}
\begin{equation}
\label{eqc2}
= \sum_{j\in [n]} \left( \sqrt{p_{g(x)}(j)p_{g(y)}(j)}
\sum_{i:x_{i}\neq y_{i}} \sqrt{p_{j,x}(i)p_{j,y}(i)} \right).
\end{equation}

As long as $g_{j}(x)\neq g_{j}(y)$,
by the definition of $p_{j}$, we have 
$\sum_{i:x_{i}\neq y_{i}} \sqrt{p_{j,x}(i)}\sqrt{p_{j,y}(i)}\ge 1/a$.
Thus we can estimate
the expression in Eq.~\ref{eqc2} from below by:

$$
\frac{1}{a} \sum_{j:g_{j}(x)\neq g_{j}(y)} \sqrt{p_{g(x)}(j)p_{g(y)}(j)}.
$$

By the definition of $p$ we can estimate
the sum (without the $1/a$ coefficient)
in the above expression from below by $1/b$,
which finishes the proof.  
\end{proof}

Another advantage of working with $\spi$ complexity is the following very
powerful lemma of Ambainis~\cite{A03} which makes it easy to lower bound the
$\spi$ complexity of iterated functions.
                                                                                
\begin{definition}
\label{df:iteration}
Let $f:\{0,1\}^n \rightarrow \{0,1\}$ be any Boolean function.  We define
the $d$th iteration of $f$, written $f^d: \{0,1\}^{n^d} \rightarrow \{0,1\}$, 
inductively as $f^1(x)=f(x)$ and
$$
f^{d+1}(x)=f(f^d(x_1, \ldots, x_{n^d}),f^d(x_{n^d+1},\ldots,x_{2n^d}), \ldots,$$
$$
f^d(x_{(n-1)n^d+1},\ldots, x_{n^{d+1}}))
$$
\end{definition}

\begin{lemma}[Ambainis] 
Let $f$ be any Boolean function and $f^d$ the $d$th
iteration of $f$.  Then $\spi(f^d)\ge \left(\spi(f)\right)^d$.
\label{le:lower_composition}
\end{lemma}

Combining this with Lemma~\ref{le:upper_composition}, we get:

\begin{corollary}
Let $f$ be any Boolean function and $f^d$ the $d$th
iteration of $f$.  Then $\spi(f^d) =  \left(\spi(f)\right)^d$.
\end{corollary}

Lemmas~\ref{le:restriction} and \ref{le:upper_composition} together with the
adversary argument lower bound
for the Grover search~\cite{Gro96,A02} imply that
for total Boolean functions, the
square root of the block sensitivity is a
lower bound on the $\spi$ complexity~\cite{A02}. Hence, by~\cite{ns94,BBCMW01}:

\begin{lemma}[Ambainis]
For total Boolean functions the $\spi$ complexity is in polynomial relation
with the various (deterministic, randomized, quantum) decision tree
complexities and the Fourier degree of the function.
\end{lemma}

\subsection{Properties of $\mpi$}

One thing that makes $\spi$ so convenient to use is that it 
dualizes~\cite{SS04}.
In this section we partially dualize the expression $\mpi$. The final
expression remains a minimization problem, but we minimize over
discrete index selection functions, instead of families of 
probability distributions, which makes it much more tractable.
Still, we remark that $\mpi$ can take exponential time (in the size of 
the truth table of $f$) whereas, $\spi$ takes polynomial time in the
size of the truth table of $f$ to compute by reduction to semidefinite
programming.

\begin{definition}[Index selection functions]
Let $f: \{0,1\}^n \rightarrow \{0,1\}$ be a Boolean function, 
$X{=}f^{-1}(0)$, and $Y{=}f^{-1}(1)$.  For $i \in [n]$ let $D_i$ be 
$|X| {\times} |Y|$ be defined by $D_i[x,y] = 1-\delta_{x_i,y_i}$. We call the 
set of $n$ Boolean ($0-1$) matrices $\{P_i\}_{i\in n}$ {\em index selection 
functions} if 
\begin{enumerate}
\addtolength{\itemsep}{-0.5\baselineskip} 
  \item{$\sum_{i} P_{i} = E$, where $E[x,y]=1$ for every $x\in X$, $y\in Y$. (informally: for every $x\in X$, $y\in Y$
we select a unique
index)}
  \item{$P_{i} \le D_{i}$ (informally: for every $x\in X$, $y\in Y$ 
the index we select is an
$i$ such that $x_{i}\neq y_{i}$).}
\end{enumerate}
\end{definition}

Notice that index selection functions correspond to partitioning
$X{\times}Y$, in such a way that if $x,y$ are in the $i$th part,
then  $x_i \neq y_i$.
\begin{theorem}[Spectral adversary version of $\mpi$]
Let $f,X,Y$ be as in the previous definition.  Let $A$ be an 
arbitrary $|X| {\times} |Y|$ nonnegative matrix satisfying $A[x,y]=0$ 
whenever $f(x)=f(y)$.  Then
$$
\label{eq:sa}
\mpi(f) = \min_{\{P_{i}\}_{i}}  
\max_A \frac{\|A\|_2}{\max_i \|A \circ P_{i}\|_2},
$$
where $\{P_{i}\}_{i}$ runs through all index selection functions.
\end{theorem}

\begin{proof}
For a fixed family of probability distributions $p=\{p_x\}$, and for the
expression
\begin{equation}
\label{eq:mpi2}
\max_{\genfrac{}{}{0pt}{}{x,y}{f(x) \not = f(y)}}
\frac{1}{\max_{i:x_{i}\neq y_{i}} \sqrt{p_x(i) p_y(i)}},
\end{equation}
\noindent let us define the index selection function $P_{i}[x,y] = 1$ if 
$i = {\rm argmax}_{i:x_{i}\neq y_{i}} \sqrt{p_x(i) p_y(i)}$ and~0
otherwise. (Argmax is the smallest argument for which the
expression attains its maximal value.)
Then the denominator in Eq.~\ref{eq:mpi2} 
becomes equal to $\sum_{i: x_i \ne y_i} \sqrt{ p_x(i) p_y(i) }
P_{i}[x,y]$.  If we replace the above system of $P_{i}$s with
any other choice of index selection function the value of $\sum_{i: x_i
\ne y_i} \sqrt{ p_x(i) p_y(i) } P_{i}[x,y]$ will not increase.
Thus we can rewrite Eq.~\ref{eq:mpi2} as
$$
\max_{\genfrac{}{}{0pt}{}{x,y}{f(x) \not = f(y)}}
\frac{1}{\max_{\{P_{i}\}_{i}}
  \sum_{i: x_i \ne y_i} \sqrt{ p_x(i) p_y(i) } P_{i}[x,y]},
$$
where here $P_{i}[x,y]$ runs through all index selection functions. 
Thus:
\begin{equation*}
\mpi(f) =
\end{equation*}
\begin{equation}1 \bigg/ \max_p \min_{\genfrac{}{}{0pt}{}{x,y}{f(x)\ne f(y)}}
 \max_{\{P_{i}\}_{i}}
  \sum_{i: x_i \ne y_i} \sqrt{ p_x(i) p_y(i) } P_{i}[x,y].
\label{eq:mpi3}
\end{equation}

Notice that in Eq.~\ref{eq:mpi3} the minimum is interchangeable with the
second maximum. The reason for this is that for a fixed $p$ there is a 
fixed $\{P_{i}[x,y]\}_{i}$ system that maximizes  
$\sum_{i: x_i \ne y_i} \sqrt{ p_x(i) p_y(i) } P_{i}[x,y]$ for all
$x,y:\; f(x)\neq f(y)$.
Thus:
\begin{equation*}\label{eq:mpi4}
\mpi(f) = 
\end{equation*}
\begin{equation*}
1 \bigg/ \max_{\{P_{i}\}_{i}}\; \max_p 
\min_{\genfrac{}{}{0pt}{}{x,y} {f(x) \ne f(y)}} 
  \sum_{i: x_i \ne y_i} \sqrt{ p_x(i) p_y(i) } P_{i}[x,y].
\end{equation*}
Following the proof of the main theorem of {\v S}palek and Szegedy we can
create the
semidefinite version of the above expression.  The difference here, however, 
is that we have to treat $\{P_{i}\}_{i}$ (the index selection functions) as 
a ``parameter'' of the semidefinite system over which we have to maximize.  
Unfortunately it also appears in the final expression.
\medskip


\noindent {\bf Semidefinite version of $\mpi$:}
For fixed $\{P_{i}\}_{i}$ let $\mu'_{\rm max}$ be
the solution of the
following semidefinite program:
\begin{equation*}
\label{eq:smm}
\begin{array}{l @{\ } r @{\ } l}
\mbox{\rm maximize} & \omit $\mu'$ \\
\mbox{\rm subject to}
  & (\forall i) \hfil R_i & \succeq 0, \\
  & \sum_i R_i \circ I &= I, \\
  & \sum_i R_i \circ P_{i} &\ge \mu' F.
\end{array}
\end{equation*}
Define $\mu_{\rm max}$ as the maximum of $\mu'_{\rm max}$, 
where $P_{i}$ ($1\le i\le n$)
run through all 
index selection functions.
Then
$\mpi = 1/\mu_{\rm max}$.

We can dualize the above program and simplify it in same way as was 
done in {\v S}palek and Szegedy for the case of $\spi$
with the only change that $D_i$ needs to be replaced with 
$P_{i}$, and that we have to minimize over all choices of $\{P_{i}\}_{i}$. 
\end{proof}

\section{Formula size lower bounds}
\label{sect:formulasize}                                                                                

\hide{
One of the most important open problems in complexity theory is proving a 
superlinear lower bound on circuit size, for an explicit function.  By 
counting, functions must exist that require exponential size, 
and most have size at least $2^n/n$~\cite{S49}, but to date the best lower 
bound known is $5n-o(n)$~\cite{LR01,IM02}.  Better bounds are known for 
formula size, where the largest lower bound for an explicit function is
$n^{3-o(1)}$~\cite{H98}.  
}

Karchmer and Wigderson \cite{KW} give an elegant characterization of formula
size in terms of a communication game.  We will use this framework to 
present our lower bounds.  This presentation has the advantage of 
showing that our methods work more generally for the communication complexity
of relations beyond the ``special case'' of formula size.  The framework of
communication complexity also allows us to work with a combinatorial 
quantity, the rectangle partition number, $C^D(R)$, which is known to lower 
bound communication complexity and arises very naturally when using $\spi$.

\subsection{Key combinatorial lemma}

We first prove a combinatorial lemma which is the key to our main result.  
This lemma relates the spectral norm squared of a matrix to the spectral 
norm squared of its submatrices.  This lemma may also 
be of independent interest.

Let $X$ and $Y$ be finite sets.  A set system $\SC$ (over $X{\times} Y$) will
be called a {\em covering} if $\cup_{S \in \SC} S = X {\times} Y$.  
Further, $\SC$ will be called a {\em partition} if $\SC$ is a covering and the
intersection of any two distinct sets from $\SC$ is empty.  
A {\em rectangle} (over $X {\times} Y$) is an arbitrary subset of 
$X {\times} Y$ of the form $X_0 {\times} Y_0$ for some $X_0 \subseteq X$ and 
$Y_0 \subseteq Y$.  A set system $\R$ will be called a 
{\em rectangle partition} if $\R$ is a partition and each $R \in \R$ is a 
rectangle.
Let $A$ be a matrix with rows indexed from $X$ and columns indexed from $Y$
and let $\R$ be a rectangle partition of $X {\times} Y$. For a rectangle 
$R = X_{0}{\times} Y_{0}\in \R$ Let $A_{R}$ be the $|X_{0}|{\times} |Y_{0}|$
submatrix of $A$ corresponding to the rectangle $R$. For 
subsets $S\subseteq X{\times} Y$ we define:
\begin{equation}\label{eq:as}
\hat A_{S}[x,y] = A[x,y], \;\; \mbox{if $(x,y)\in S$ and $0$ otherwise}.
\end{equation}
Notice that for a rectangle $R$, matrices $A_{R}$ 
and $\hat A_{R}$ differ only by
a set of all-zero rows and columns. We are now ready to state the lemma:
\begin{lemma}
Let $A$ be an arbitrary $|X| {\times} |Y|$ matrix (possibly with complex entries),
and $\R$ a partition of $X {\times} Y$.  Then
$\|A\|_2^2 \le \sum_{R \in \R} \|A_R\|_2^2$
\label{le:pretty_lemma}
\end{lemma}

\begin{proof}
By Proposition~\ref{pr:fact}, $\|A\|_2=\max_{u,v} |u^*Av|$, where the maximum
is taken over all unit vectors $u,v$.  Let $u,v$ be the unit vectors 
realizing this maximimum.  Then we have
$$
\|A\|_2= |u^*Av| = \left|u^* \left(\sum_{R \in \R} \hat A_R \right) v \right| =
\left|\sum_{R \in \R} u^* \hat A_R v \right|.
$$
Let $u_R^*$ be the portion of $u^*$ corresponding to the rows of $R$, and
$v_R$ be the portion of $v$ corresponding to the columns of $R$.  Notice that
$\{u_R\}_{R \in \R}$ do not in general form a partition of $u$.
We now have
\begin{eqnarray*}
\|A\|_2 &=& \left|\sum_{R \in \R} u_R^* A_R v_R \right| \le
\sum_{R \in \R} \left| u_R^* A_R v_R \right| \\
&\le& \sum_{R \in \R} \|A_R\|_2 |u_R| |v_R|
\end{eqnarray*}
by Proposition~\ref{pr:fact}.  Applying the Cauchy--Schwarz inequality, we
obtain
$$
\|A\|_2 \le 
\left( \sum_{R \in \R} \|A_R\|_2^2 \right)^{1/2}
        \left( \sum_{R \in \R} |u_R|^2 |v_R|^2 \right)^{1/2}.
$$
Now it simply remains to observe that
$$
\sum_{R \in \R} |u_R|^2 |v_R|^2 = \sum_{R \in \R}
\sum_{(x,y) \in R} u[x]^2 v[y]^2= |u|^2 |v|^2 =1,
$$
as $\R$ is a partition of $X {\times} Y$.
\end{proof}

\subsection{Deterministic formulae}

In this section, we prove our main result that 
$\mpi$ is a lower bound on formula size.
We first identify two natural properties which
are sufficient for a function to be a formula 
size lower bound.  

\begin{definition}
A function $\mu:2^{X{\times} Y}\rightarrow \Reals^{+}$
is called a {\em rectangle measure} if the following properties hold. 
\begin{enumerate}
\addtolength{\itemsep}{-0.5\baselineskip} 
\item (Subadditivity) For any rectangle partition
$\R$ of $X{\times} Y$, $\mu(X{\times} Y) \leq \sum_{R\in \R} \mu(R)$.
\item (Monotonicity) For any rectangle $R \subseteq X{\times} Y$, and
subset $S \subseteq X{\times} Y$, if $R\subseteq S$ then
$\mu(R) \leq \mu(S)$.
\end{enumerate}
\end{definition}

Theorem \ref{le:pretty_lemma} implies that for any $|X|{\times} |Y|$ matrix 
$A$ with non-negative entries $S\rightarrow  ||\hat A_{S}||$ of 
expression (\ref{eq:as}) is a rectangle measure.
Other examples include the rank of $\hat A_{S}$ for any matrix $A$
over any field (see Section~\ref{sect:Razborov}), and
the $\mu$-rectangle size bounds of~\cite{KKN95}.

Let ${\SC_1, \SC_2}$ be two families of sets over the same universe. 
We say that ${\SC}_1$ is {\em embedded} in ${\SC}_2$ ($\SC_1 \embedded \SC_2$) 
if for every 
$S \in {\SC}_1$ there is a $S' \in {\SC}_2$ such that 
$S \subseteq S'$.  

\begin{theorem}
\label{thm:mu}
Let $\mu$ be a rectangle measure over $2^{X{\times} Y}$, $\SC$
be a covering of $X{\times} Y$ and $\R$ a rectangle
partition of $X {\times} Y$ such that $\R \embedded \SC$.
Then $|\R| \geq \frac{\mu(X{\times} Y)}{\max_{S\in\SC}\mu(S) }$.
\end{theorem}
The proof follows by subadditivity and monotonicity of $\mu$.

\begin{theorem}[Main Theorem]
\[
\spi^2(f) \le \mpi^2(f) \le C^D(R_f) \le \L(f)
\]
\label{thm:main}
\end{theorem}

\begin{proof}
We have seen that $\spi^2(f) \le \mpi^2(f)$, and $C^D(R_f) \le \L(f)$
follows from the Karchmer--Wigderson communication game characterization of
formula size, thus we focus on the inequality $\mpi^2(f) \le C^D(R_f)$.  

Let $\R$ be a monochromatic rectangle partition of $R_f$ such that 
$|\R|=C^D(R_f)$, and let $A$ be an arbitrary $|X| {\times} |Y|$ matrix with
nonnegative real entries.  For $R \in \R$ let $\mbox{ color}(R)$ be 
the least index $c$ such that $x_c \ne y_c$ holds for all $(x,y)\in R$.  By 
assumption each $R$ is monochromatic, thus such a color exists. Define
\[
S_c=\cup_{\mbox{ color}(R)= c} R.
\]
Then $\R$ is naturally embedded in the covering $\{S_c\}_{c \in [n]}$.  
For any $S\subseteq X{\times} Y$, let $\mu_A(S) =  \| \hat A_{S}\|_2^2$.
By Lemma~\ref{le:pretty_lemma}, and item 3 of Proposition~\ref{pr:fact}, 
$\mu_A$ is a rectangle measure.  Hence by Theorem~\ref{thm:mu},
$$
\max_A \frac{\|A\|_2^2}{\max_c \| \hat A_{S_c}\|_2^2} \le C^D(R_f).
$$
We have exhibited a particular index selection function, the $\{S_c\}_c$, 
for which this inequality holds, thus it also holds for $\mpi^2(f)$ which is
the minimum over all index selection functions.
\end{proof}

\subsection{Probabilistic Formulae}
The properties of $\spi$ allow us to show that it can be used
to lower bound the probabilistic formula size.

\begin{lemma}
\label{Le:spi_rand}
Let $\epsilon < 1/2$.
If $f:S \rightarrow \{0,1\}$ is $\epsilon$-approximated by functions
$\{f_j\}_{j \in J}$ with $\spi(f_j) \le s$ for every $j \in J$, then
$\spi(f) \le s/(1-2\epsilon)$.
\end{lemma}
                                                                                
\begin{proof}
By assumption there is a probability distribution 
$\alpha=\{\alpha_j\}_{j\in J}$ such that $\Pr[f(x)=f_j(x)]\ge 1- \epsilon$.
Thus for a fixed $x \in S$, letting $J_x=\{j \in J: f(x)=f_j(x)\}$, we have
$\sum_{j \in J_x} \alpha_j \ge 1-\epsilon$.  Hence for any $x,y \in S$
we have $\sum_{j \in J_x \cap J_y} \alpha_j \ge 1-2\epsilon$.
For convenience, we write $J_{x,y}$ for $J_x \cap J_y$.
As $\spi(f_j) \le s$ there is a family of probability
distributions $p_j$ such that
whenever $f_j(x)\not= f_j(y)$
$$
\Isum \sqrt{p_{j,x}(i)p_{j,y}(i)}\ge 1/s.
$$
Define $p_x(i)=\sum_{j\in J} \alpha_j p_{j,x}(i)$.  Let $x,y$ be such that
$f(x)\not = f(y)$.  
\begin{eqnarray*}
\lefteqn{\Isum \sqrt{p_x(i)p_y(i)}}\\&=&\Isum \sqrt{\sum_{j\in J} \alpha_j p_{j,x}(i)}
        \sqrt{\sum_{j \in J}\alpha_j p_{j,y}(i))} \\
&\ge& \Isum \sqrt{\sum_{j\in J_{x,y}} 
        \alpha_j p_{j,x}(i)} \sqrt{\sum_{j\in J_{x,y}}
        \alpha_j p_{j,y}(i)} \\
&\ge& \Isum \sum_{j \in J_{x,y}} \sqrt{\alpha_j p_{j,x}(i)}
        \sqrt{\alpha_j p_{j,y}(i)} \\
&=& \sum_{j \in J_{x,y}} \left(\alpha_j \Isum \sqrt{p_{j,x}(i)p_{j,y}(i)} \right)\\
&\ge& \frac{1-2\epsilon}{s},
\end{eqnarray*}
where for the third step we have used the Cauchy--Schwarz Inequality.
\end{proof}
                                                                                
This lemma immediately shows that the $\spi$ method can give lower bounds
on probabilistic formula size.
\begin{theorem}
Let $S\subseteq \{0,1\}^n$ and $f:S\rightarrow \{0,1\}$.  Then
$\L^\epsilon(f) \ge \left((1-2\epsilon) \spi(f) \right)^2$ for any
$\epsilon < 1/2$.
\end{theorem}
                                                                                
\begin{proof}
Suppose that $\{f_j\}_{j \in J}$ gives an $\epsilon$-approximation to $f$.
Using Lemma~\ref{Le:spi_rand} in the contrapositive implies that there exists
some $j \in J$ with $\spi(f_j)\ge (1-2\epsilon)\spi(f)$.
Theorem~\ref{thm:main} then implies
$\L(f_j)\ge \left( (1-2\epsilon)\spi(f) \right)^2$ which gives the
statement of the theorem.
\end{proof}
                                                                                
\section{Comparison among methods}
\label{sect:comparison}
In this section we look at several formula size lower bound techniques 
and see how they compare with our methods.  A bottleneck in
formula size lower bounds seems to have been to go beyond methods which only
consider pairs $(x,y)$ with $f(x)\ne f(y)$ which have Hamming distance 1.  
In fact, the methods of Khrapchenko, Koutsoupias, and a lemma of H{\aa}stad can
all be seen as special cases of the $\spi$ method where only pairs of Hamming
distance 1 are considered.

\subsection{Khrapchenko's method }
One of the oldest and most general techniques available for showing formula 
size lower bounds is Khrapchenko's method \cite{K71}, originally used to give 
a tight $\Omega(n^2)$ lower bound for the parity function.  This method 
considers a bipartite graph whose left vertices are the 0-inputs to $f$ and 
whose right vertices are the 1-inputs.  The bound given is the product of 
the average degree of the right and left hand sides.
\begin{theorem}[Khrapchenko]
\label{th:Khrap}
Let $S \subseteq \{0,1\}^n$ and $f:S \rightarrow \{0,1\}$.  Let 
$A \subseteq f^{-1}(0)$ and $B \subseteq f^{-1}(1)$.  Let 
$C$ be the set of pairs $(x,y) \in A {\times} B$ with Hamming distance 1,  
that is $C=\{(x,y) \in A {\times} B : d_H(x,y)=1\}$.  
Then $\L(f) \ge \spi(f)^2 \ge \frac{|C|^2}{|A||B|}$.
\end{theorem}

Khrapchenko's method can easily be seen as a special case of the
probability scheme.  Letting $A,B,C$ be as in the statement of the theorem, 
we set up our probability distributions as follows:
\begin{myitemize}
  \item{$p_A(x){=}1/|A|$ for all $x {\in} A$, $p_A(x){=}0$ otherwise}
  \item{$p_B(x){=}1/|B|$ for all $x {\in} B$, $p_B(x){=}0$ otherwise}
  \item{$q(x,y){=}1/|C|$ for all $(x,y) {\in} C$, $q(x,y){=}0$ otherwise}
  \item{$p_{x,i}(y){=}1$ if $(x,y){\in} C$ and $x_i \not = y_i$, 0 otherwise.
Note that this is a probability distribution as for every $x$ there is only
one $y$ such that $(x,y) {\in} C$ and $x_i \neq y_i$.}
\end{myitemize}
By Theorem~\ref{th:SS04} and Theorem~\ref{thm:main},
$$
\L(f) \ge \min_{\genfrac{}{}{0pt}{}{x,y,i}
                                   {
\genfrac{}{}{0pt}{}{
             f(x)\not = f(y),}{x_i \not = y_i}
                                   }}
\frac{p_A(x)p_B(y)p_{x,i}'(y)p_{y,i}'(x)}{q(x,y)}
= \frac{|C|^2}{|A||B|},
$$
where the expression in the middle is a lower bound on $\spi(f)^2$.

\subsection{The Koutsoupias bound}
Koutsoupias~\cite{Kou93} extends Khrapchenko's method with
a spectral version.  The weights are always 1 for pairs of inputs
with different function values that have Hamming distance 1, and~0
everywhere else.
                                                                                
\begin{theorem}[Koutsoupias]
Let $f:\{0,1\}^n\rightarrow\{0,1\}$, and let $A\subseteq f^{-1}(0)$,
and $B\subseteq f^{-1}(1)$. Let $C=\{(x,y)\in A{\times} B : d_H(x,y)=1\}$.
Let $Q$ be a $|B|{\times} |A|$ matrix $Q[x,y]=C(x,y)$ where $C$
is identified with its characteristic function.
Then $\L(f) \geq \spi(f)^2 \geq \|Q\|_2^2$. 
\label{th:Kou}
\end{theorem}
                                                                                
\begin{proof}
The bound follows easily from the the spectral version of $\spi$.  Let
$Q$ be as in the statement of the theorem.  Notice that since we only
consider pairs with Hamming distance 1, for every row and column of 
$Q_i$ there is at most one nonzero entry, which is at most 1.  Thus 
by Proposition~\ref{pr:fact} we have 
$\|Q_i\|_2^2 \le \|Q\|_1 \|Q\|_{\infty} \le 1$.  The theorem now follows
from Theorem~\ref{thm:main}.
\end{proof}

\subsection{H{\aa}stad's method}
                                                                                
The shrinkage exponent of Boolean formulae is the
least upper bound $\gamma$ such that subject to a random
restriction where each variable is left free with probability $p$,
Boolean formulae shrink from size $L$ to expected size $p^\gamma L$.
Determining the shrinkage exponent is important as 
Andreev \cite{And87} defined a function $f$ whose formula size 
is $\L(f)=n^{1+\gamma}$.  H{\aa}stad \cite{H98} shows the shrinkage
exponent of Boolean formulae is 2 and thereby obtains an $n^3$ formula size
lower bound (up to logarithmic factors), the largest bound known for an
explicit function.  On the way to this result, H{\aa}stad proves an 
intermediate lemma which gives a lower bound on formula size that depends on 
the probability that restrictions of a certain form occur.  He proves that 
this lemma is a generalization of Khrapchenko's method; we prove that 
H{\aa}stad's lemma is in turn a special case of $\spi$.  Since H{\aa}stad's 
method uses random restrictions, which at first glance seems completely
different from adversary methods, it comes as a surprise that
it is in fact a special case of our techniques.

\begin{definition}
For any function $f:\{0,1\}^n\rightarrow \{0,1\}$,
\begin{enumerate}
\addtolength{\itemsep}{-0.5\baselineskip} 
\item A {\em restriction} is a string in $\{0,1,\star\}^n$ 
where $\star$ means the
variable is left free, and~0 or~1 mean the variable
is set to the constant~0 or~1, respectively.
\item The restricted function $f|_\rho$ is the 
function that remains after the non-$\star$ variables in $\rho$
are fixed.
\item $R_p$ is the distribution on random restrictions to the variables of
$f$ obtained by setting each variable, independently, to $\star$
with probability $p$, and to~0 or~1 each with probability $\frac{(1-p)}{2}$.
\item A {\em filter} $\Delta$ is a set of restrictions which has
the property that if $\rho\in \Delta$, then every $\rho'$
obtained by fixing one of the $\star$s to a constant is also
in $\Delta$.
\item When $p$ is known from the context, and for any event $E$, and any filter $\Delta$,
we write $\Pr[E|\Delta]$ to mean $\Pr_{\rho\in R_p}[E|\rho\in \Delta]$.
\end{enumerate}
\end{definition}
                                                                                
\begin{theorem}[H{\aa}stad, Lemma~4.1]
\label{thm:hastad}
Let $f:\{0,1\}^n\rightarrow\{0,1\}$.
Let $A$ be the event
that a random restriction in $R_p$
reduces $f$ to the constant~0,
$B$ be the event
that a random restriction in $R_p$
reduces $f$ to the constant 1,
and let $C$ be the event that a random restriction $\rho\in R_p$
is such that $f|_\rho$ is a single literal.
Then $$ \L(f) \geq \frac{Pr[C|\Delta]^2}{Pr[A|\Delta]Pr[B|\Delta]}
\left(\frac{1-p}{2p}\right)^2$$
\end{theorem}
\begin{proof}
We show that the theorem follows from the probability scheme
(Definition~\ref{df:prob_scheme}).
In this proof we only consider restrictions obtained from $R_p$
that are in the filter $\Delta$.
We also abuse notation and use $A$ and $B$ to mean the sets 
of restrictions in $\Delta$
which contribute with non-zero probability to the events 
$A$ and $B$ respectively.
                                                                                
Implicit in H{\aa}stad's proof is the following relation between
restrictions in $A$ and $B$.  For every $\rho \in C$, $f|_\rho$
reduces to  a single literal, that is, for every $\rho \in C$, there 
is an $i$ such that $f|_\rho = x_i$ (or $\neg x_i$ if the variable 
is negated).  Define $\rho^b$ to be $\rho$
where $x_i$ is set to $b$, for $b\in \{0,1\}$  (set $x_i$
to $1{-}b$ if the variable is negated).
To fit into the framework of the probability scheme,
let $\overline{\rho^b}$ be $\rho^b$ where
all remaining $\star$s are set to~1.  This doesn't change
the value of the function, because it is already constant on $\rho^b$.
Then we say that $\overline{\rho^0}, \overline{\rho^1}$ are in the relation.
                                                                                
We set
$p_A(\sigma) = \frac{\Pr[\sigma]}{\Pr[A|\Delta]}$ for any $\sigma \in A$, and
$p_B(\tau) = \frac{\Pr[\tau]}{\Pr[B|\Delta]}$ for any $\tau \in B$, 
and for every pair $\overline{\rho^0}, \overline{\rho^1}$  in the relation,
where  $\rho \in C$, $f|_\rho = x_i$ or $\neg x_i$, let
\begin{eqnarray*}
p'_{\overline{\rho^0},i}(\overline{\rho^1}) &=&1\\
p'_{\overline{\rho^1},i}(\overline{\rho^0}) &=&1\\
q(\overline{\rho^0},\overline{\rho^1}) &=& \frac{Pr[\rho]}{Pr[C|\Delta]}
\end{eqnarray*}
The probabilities are~0 on all other inputs.
We can easily verify that the probabilities sum to~1.
For $p'$, notice
that the Hamming distance between
$\overline{\rho^0}$ and $\overline{\rho^1}$ is 1, so
when $\overline{\rho^b}$ and $i$ are fixed, 
there is only a single 
$\overline{\rho^{1-b}}$ with probability 1.

By Theorem~\ref{th:SS04} and Theorem~\ref{thm:main},
\begin{eqnarray*}
\L(f) &\geq&  \frac{p_A(x)p_B(y)p'_{y,i}(x)p'_{x,i}(y)}{q(x,y)^2}\\
  & = & {\frac{Pr[\rho^0]}{Pr[A|\Delta]}\frac{Pr[\rho^1]}{Pr[B|\Delta]}}
     {\left(\frac{Pr[C|\Delta]} {Pr[\rho]}\right)^2}\\
\end{eqnarray*}
Finally, notice that 
$Pr[\rho]=\frac{2p}{1-p}Pr[\rho^b]$.
\end{proof}

\noindent{\bf Remark} 
H{\aa}stad actually defines $f|_\rho$ to be the result of 
reducing the formula for $f$ (not the function)
by applying a sequence of reduction rules,
for each restricted variable.  
So there is a subtlety here about whether 
$f|_\rho$ denotes  the reduced formula, 
or the reduced function, and the probabilities might
be different if we are in one setting or the other.  
However both in his proof and ours, the only thing that
is used about the reduction is that if the formula or
function reduces to a single literal, then fixing this literal 
to 0 or to 1 reduces the function to a constant.
Therefore, both proofs go through for both settings.

\subsection{Razborov's method}
\label{sect:Razborov}
Razborov 
\cite{R90} proposes a formula size lower bound technique using matrix rank:
\begin{theorem}[Razborov]
Let $\SC$ be a covering 
over $X {\times} Y$, let 
$A$ be an arbitrary nonzero $|X| {\times} |Y|$ 
matrix, and
$\R$ be a rectangle partition of $X {\times} Y$ such 
that $\R \embedded \SC$.
Then
$$
\max_A \frac{\rk(A)}{\max_{S \in \SC} 
\rk(\hat A_S)} \le \alpha(\SC).
$$
\label{thm:razborov}
\end{theorem}
It can be easily verified that the function $S \rightarrow \rk(\hat A_S)$ is 
a rectangle measure, thus this theorem follows from Theorem~\ref{thm:mu}.
Razborov uses Theorem~\ref{thm:razborov} to show superpolynomial monotone 
formula size lower bounds, but also shows that the method becomes trivial 
(limited to $O(n)$ bounds) for regular formula size 
\cite{R92}.  An interesting difference between matrix rank and 
and spectral norm is that $\rk(A+B) \le \rk(A) + \rk(B)$ holds 
for any two matrices $A,B$, while a necessary condition for 
subadditivity of the spectral norm squared is that $A,B$ be disjoint 
rectangles.

\section{Limitations}
\label{sect:limits}

\subsection{Hamming distance 1 techniques}
We show that the bounds for a function $f$ given by Khrapchenko's and 
Koutsoupias' method, and by H{\aa}stad's lemma are upper bounded by the 
product of the zero sensitivity and the one sensitivity of $f$.  We will later 
use this bound to show a function on $n$ bits for which the best lower bound
given by these methods is $n$ and for which an $n^{1.32}$ bound
is provable by $\spi^2$.  
 
\begin{lemma}
\label{le:Kh-limits}
The bound given by the Khrapchenko method (Theorem~\ref{th:Khrap}), 
Koutsoupias' method (Theorem~\ref{th:Kou}), and H{\aa}stad's Lemma 
(Theorem~\ref{thm:hastad}) for a function $f$ are at 
most $s_0(f)s_1(f) \le s^2(f)$.
\end{lemma}

\begin{proof}
Let $A$ be a nonnegative matrix, with nonzero entries only in positions
$(x,y)$ where $f(x)=0, f(y)=1$ and the Hamming distance between $x,y$ is one.
We first show that  
\begin{equation}
\max_A \frac{\|A\|_2^2}{\max_i \|A_i\|_2^2} \le s_0(f) s_1(f).
\label{eq:h1_sp}
\end{equation}
Let $a_{max}$ be the largest entry in $A$.  $A$ can have at most $s_0(f)$ many
nonzero entries in any row, and at most $s_1(f)$ many nonzero entries in 
any column, thus by item 2 of Propostion~\ref{pr:fact}, 
$$
\|A\|_2^2 \le \|A\|_1 \|A\|_{\infty} \le a_{max}^2 s_0(f) s_1(f).
$$   
On the other hand, for some $i$, the entry $a_{max}$ appears in $A_i$, and so
by item 1 of Proposition~\ref{pr:fact}, $\|A_i\|_2^2 \ge a_{max}^2.$  
Equation~\ref{eq:h1_sp} follows. 

Now we see that the left hand side of Equation~\ref{eq:h1_sp} is larger than
the three methods in the statement of the theorem.  That it is more general
than Koutsoupias method is clear.  To see that it is more general than the
probability schemes method where $q(x,y)$ is only positive if the Hamming
distance between $x,y$ is one: given the  probability distributions
$q,p_X,p_Y$, define the matrix 
$A[x,y]=q(x,y)/\sqrt{p_X(x)p_Y(y)}$.  By item 1 of Proposition~\ref{pr:fact}, 
$\|A\|_2 \ge 1$, witnessed by the unit vectors $u[x]=\sqrt{p_X(x)}$ and 
$v[y]=\sqrt{p_Y(y)}$.  As each reduced matrix $A_i$ has at most one nonzero 
entry in each row and column, by item 2 of Proposition~\ref{pr:fact} we have
$$
\max_i \|A_i\|_2^2 \le \max_{x,y} \frac{q^2(x,y)}{p_X(x)p_Y(y)}.
$$
Thus we have shown
$$
\max_A \frac{\|A\|_2^2}{\max_i \|A_i\|_2^2} \ge \max_{p_X,p_Y,q} \min_{x,y}
\frac{p_X(x)p_Y(y)}{q^2(x,y)}.
$$
\end{proof}

The only reference to the limitations of these methods we are aware of is
Sch\"urfeld~\cite{Sch86}, who shows that Khrapchenko's method cannot prove 
bounds greater than $C_0(f) C_1(f)$.

\subsection{Limitations of $\spi$ and $\mpi$}
The limitations of the adversary method are well known
\cite{A02,LM04,Sze03,Zha03,SS04}.
\v{S}palek and Szegedy, in unifying the adversary methods, also give the
most elegant proof of their collective limitation.  The same proof also shows
the same limitations hold for the $\mpi$ measure.
\begin{lemma}
\label{le:mpi_limit}
Let $f:\{0,1\}^n \rightarrow \{0,1\}$ be any partial or total Boolean
function.  If $f$ is total (respectively, partial) then
$\mpi(f)\le \sqrt{C_0(f)C_1(f)}$ (respectively, 
$\min\{\sqrt{nC_0(f)},\sqrt{nC_1(f)}\}$).
\end{lemma}
                                                                                
\begin{proof}
Assume that $f$ is total.  Take $x,y$ such that $f(x)=0$ and $f(y)=1$.
We choose any 0-certificate $B_0$ for $x$ and any 1-certificate $B_1$ for $y$
and let $p_x(i)=1/C_0(f)$ for all $i \in B_0$ and $p_y(i)=1/C_1(f)$ for all
$i \in B_1$.  As $f$ is total, we have $B_0 \cap B_1 \not = \emptyset$, thus let $j \in B_0 \cap B_1$.  For this $j$ we have
$p_x(j)p_y(j)\ge 1/\left(C_0(f)C_1(f)\right)$, thus
$\min_i 1/p_x(i)p_y(i) \ge C_0(f)C_1(f)$.
                                                                                
The case where $f$ is partial follows similarly.  As we no longer know
that $B_0 \cap B_1 \not = \emptyset$, we put a uniform distribution over
a 0-certificate of $x$ and the uniform distribution over $[n]$ on $y$ or
vice versa.
\end{proof}
                                                                                
This lemma implies that $\spi$ and $\mpi$ are polynomially related for total
$f$.

\begin{corollary}
Let $f$ be a total Boolean function.  Then $\mpi(f) \le \spi^4(f)$.
\end{corollary}

\begin{proof}
By \cite[Thm. 5.2]{A02} we know that $\sqrt{bs(f)} \le \spi(f)$.  As 
$f$ is total, by the above lemma we know that 
$\mpi(f) \le \sqrt{C_0(f) C_1(f)}$.  This in turn is smaller than 
$bs(f)^2$ as $C(f) \le s(f) bs(f)$ \cite{N91}.  The statement 
follows.
\end{proof}

Besides the certificate complexity barrier, another serious limitation of the 
$\spi$ method occurs for partial functions where
every positive input is far in Hamming distance from every negative input.
Thus for example, if for any pair $x,y$ where $f(x)=1$ and $f(y)=0$ we have
$d_H(x,y)\ge \epsilon n$, then by putting the uniform distribution over
all input bits it follows that $\spi(f) \le 1/\epsilon$.  The measure $\mpi$
does not face this limitation as there we still only have one term in the 
denominator.

Following this line of thinking, we can give an example of a partial function
$f$ where $\mpi(f) \gg \spi(f)$.  Such an example is the Collision problem
(see Section~\ref{sect:collision}), as here any positive and negative 
input must differ on at least $n/2$ positions.   
Another family of examples comes from property testing,
where the promise is that the input either has some property,
or that it is $\epsilon$-far from having the property.

\begin{figure}
\begin{center}
\epsfig{file=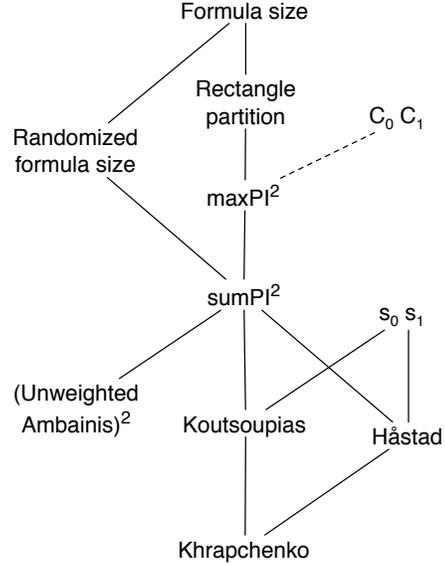,width=6cm} 
\end{center}
\caption{\rm{Summary of the methods and their limitations.
The containments denoted by solid lines hold for total as well as partial
functions.  All containments are strict.}}
\end{figure}

\section{Concrete lower bounds}
\label{sect:applications}

The quantum adversary argument has been used to prove
lower bounds for a variety of problems.  Naturally,
all of these lower bounds carry over to formula size
lower bounds.  In this section we present some new lower bounds,
in order to highlight the strengths and weaknesses of
$\mpi$ and $\spi$.

\subsection{Recursive majorities}
As an example of applying $\spi$, we look at the recursive majority
of three function.  We let $\MAJ{h} : \{0,1\}^{3^h}\rightarrow \{0,1\}$ be
the function computed by a complete ternary tree of depth $h$ where every
internal node is labeled by a majority gate and the input is given at the
leaves.
                                                                                
Recursive majority of three has been studied before in various contexts.
It is a monotone function which is very sensitive to noise~\cite{MO02},
making it useful for hardness amplification in NP~\cite{OD02}.
Jayram, Kumar, and Sivakumar~\cite{JKS03} give nontrivial lower and upper
bounds on the randomized decision tree complexity of recursive majority of
three.  They show a lower bound of $(7/3)^h$ on the randomized
decision tree complexity.  As far as we know, the quantum query complexity
of recursive majority of three has not yet been investigated.  We show a lower
bound of $2^h$ on the quantum query complexity.
                                                                                
\begin{lemma}
$\spi(\MAJ{h})=\mpi(\MAJ{h})=2^h$
\label{le:spi-rec-maj}
\end{lemma}
                                                                                
\begin{proof}
To see that $\mpi(\MAJ{h}) \le 2^h$, observe that
$C_0(\MAJ{h})=C_1(\MAJ{h})=2^h$.  The result then follows from
Lemma~\ref{le:mpi_limit}.
                                                                                
We now turn to the lower bound.  We will first show a lower bound for
$\MAJ{1}$, the majority of three function, and
then apply Lemma~\ref{le:lower_composition}.  Consider the following table,
where the rows are indexed by negative instances $x$, the columns by positive
instances $y$, and 1's indicate when $d_H(x,y)=1$.
\begin{center}
\begin{tabular}{c|ccc}
     & 110 & 101 & 011  \\ \hline
001  &  0  &  1  &  1   \\
010  &  1  &  0  &  1   \\
100  &  1  &  1  &  0  \\
\end{tabular}
\end{center}
Interpreting this table as the adjacency matrix of a graph, it
is clear that every vertex has degree 2.  Thus Khrapchenko's method
gives a bound of 4 for the base function.  By Theorem~\ref{th:Khrap}
we have $\spi(\MAJ{1})\ge 2$.  Now applying Lemma~\ref{le:lower_composition}
gives the lemma.
\end{proof}
                                                                                
From Lemma~\ref{le:spi-rec-maj} we immediately obtain quantum query
complexity and formula size lower bounds:
\begin{theorem}
\label{th:lower-rec-maj}
Let $\MAJ{h}$ be the recursive majority of three function of height $h$.  
Then $Q_\epsilon(\MAJ{h}) \ge (1-2\sqrt{\epsilon (1-\epsilon)})2^h$ and
$\L^\epsilon(\MAJ{h}) \ge (1-2\epsilon)4^h$.
\end{theorem}
                                                                                
The best upper bound on the formula size of $\MAJ{h}$ is $5^h$.
For this bound, we will use the
following simple proposition about the formula size of iterated functions.
\begin{proposition}
\label{pr:formula_size_iteration}
Let $S \subseteq \{0,1\}^n$ and $f: S {\rightarrow} \{0,1\}$.  If $\L(f) \le s$
then $\L(f^d) \le s^d$, where $f^d$ is the $d^{th}$ iteration of $f$.
\end{proposition}

\begin{proposition}
$\L(\MAJ{h}) \le 5^h$.
\end{proposition}
                                                                        
\begin{proof}
The formula $(x_1 \wedge x_2) \vee ((x_1 \vee x_2) \wedge x_3)$ computes
$\MAJ{1}$ and has 5 leaves.  Using Proposition~\ref{pr:formula_size_iteration} 
gives $\L(\MAJ{h}) \le 5^h$.
\end{proof}

\subsection{Ambainis' function}
We define a function $\A:\{0,1\}^4 \rightarrow \{0,1\}$ after
Ambainis~\cite{A03}.  This function evaluates to 1 on the following
values: 0000, 0001, 0011, 0111, 1111, 1110, 1100, 1000.
That is, $f(x) = 1 $ when $x_1\leq x_2 \leq x_3 \leq x_4 $ or 
$x_1\geq x_2 \geq x_3 \geq x_4$.  To obtain this formulation from Ambainis' 
original definition, exchange $x_1$ and $x_3$, and take the negation of the 
resulting function.  There are a few things to notice
about this function.  The sensitivity of $\A$ is 2 on every input.  Also on 
an input $x=x_1 x_2 x_3 x_4$ the value of $\A(x)$ changes if both bits 
sensitive to $x$ are flipped simultaneously, and if both bits insensitive for $x$ are 
flipped simultaneously.

We will be looking at iterations of the base function $\A$ as in 
Definition~\ref{df:iteration}.  Notice that the sensitivity of $\A^d$ is 
$2^d$ on every input $x \in \{0,1\}^{4^d} $. 
                                                                                
\begin{lemma}
$\spi(\A^d)=2.5^d$.
\end{lemma}

\begin{proofnobox}
Ambainis has already shown that $\spi(\A^d)\ge 2.5^d$~\cite{A03}.
                                                                                
We now show the upper bound.  We will show an upper bound for the base
function $\A$ and then use the composition Lemma~\ref{le:upper_composition}.
Every input $x_1x_2x_3x_4$ has two sensitive variables and two insensitive
variables.  For any $x \in \{0,1\}^4$ we set $p_x(i)=2/5$ if $i$ is
sensitive for $x$ and $p_x(i)=1/10$ if $i$ is insensitive for $x$.
The claim follows from the following observation: for any $x,y \in \{0,1\}^4$
such that $f(x)\not =f(y)$ at least one of the following holds
\begin{myitemize}
  \item{$x$ and $y$ differ on a position $i$ which is sensitive for both
$x$ and $y$.  Thus $\sum_i\sqrt{p_x(i)p_y(i)}\ge 2/5$}
  \item{$x$ and $y$ differ on at least 2 positions, each of these positions
being sensitive for at least one of $x,y$.  Thus
$\sum_i\sqrt{p_x(i)p_y(i)}\ge 2\sqrt{1/25}=2/5$}\hfill $\Box$
\end{myitemize}
\nopagebreak
\end{proofnobox}
                                                                                
This lemma gives us a bound of $6.25^d \approx N^{1.32}$ on the formula size of 
${\A}^d$.  Since the sensitivity of 
${\A}^d$ is 
$2^d$, by Lemma~\ref{le:Kh-limits}, 
the best bound provable by Khrapchenko's method, Koutsoupias' method,
and H{\aa}stad's lemma is $4^d=N$.

It is natural to ask how tight this formula size bound is.  The best upper 
bound we can show on the formula size of $\A^d$ is $10^d$.  

\begin{proposition}
$\L(\A^d) \le 10^d$
\end{proposition}

\begin{proof}
It can be easily verified that the following formula of size 10 computes the 
base function $\A$: 
\begin{eqnarray*}
\lefteqn{(\neg x_1 \vee x_3 \vee \neg x_4) \wedge }\\
&&\left( 
(\neg x_1 \wedge x_3 \wedge x_4) \vee \left( (x_1 \vee \neg x_2) \wedge 
(x_2 \vee \neg x_3) \right) \right).
\end{eqnarray*}
This formula was found by computer search.
The claim now follows from Proposition~\ref{pr:formula_size_iteration}.  
\end{proof}
                                                                                
\subsection{Collision problem}
\label{sect:collision}
In this section we look at the collision problem.  This is a promise
problem, where for an alphabet $\Sigma$ the inputs $x=x_1x_2\ldots x_n \in
\Sigma^n$ satisfy one of the
following conditions:
\begin{myitemize}
  \item{All $x_i$ are different}
  \item{For each $i$ there exists exactly one $j\not =i$ such that $x_i=x_j$.}
\end{myitemize}
Those inputs satisfying the first condition are positive inputs and those
satisfying the second condition are negative.  An optimal lower bound
for the quantum query complexity of $\Omega(n^{1/3})$ has been given by
Aaronson and Shi~\cite{AY04}.  We now show that the quantum adversary
method cannot give better than a constant bound for this problem.
\begin{lemma}
$\spi(\Col) \le 2$
\end{lemma}
                                                                                
\begin{proof}
We demonstrate a set of probability distributions $p_x, p_y$ such that for any
positive instance $x$ and negative instance $y$ we have
$$
\Isum \sqrt{p_x(i)} \sqrt{p_y(i)} \ge 1/2.
$$
The upper bound then follows.
                                                                                
Our probability distribution is simple: for every $x$, let $p_x(i)$ be
the uniform distribution over $[n]$.  Any positive and negative instance
must disagree in at least $n/2$ positions, so
$$
\Isum \sqrt{p_x(i)}\sqrt{p_y(i)} \ge \frac{n}{2} \sqrt{\frac{1}{n}\frac{1}{n}}
=\frac{1}{2}.
$$ \nopagebreak \end{proof}
                                                                                
On the other hand, $\mpi(\Col) \ge \sqrt{n/2}$.  As there is an upper bound
for the collision problem of $O(n^{1/3})$ by Brassard, H{\o}yer, Tapp~\cite{BHT97},
this also shows that in general $\mpi(f)$ is not a lower bound on the
quantum query complexity of $f$.
                                                                                
\begin{lemma}
$\mpi(\Col) = \Theta(\sqrt{n})$
\end{lemma}
                                                                                
\begin{proof}
For the upper bound: On every positive instance $x$, where all $x_i$ are
different, we put the uniform distribution over $i \in [n]$; for a negative
instance $y$ we put probability $1/2$ on the first position, and probability 
$1/2$ on the position $j$ such that $y_1=y_j$.  As $y_1=y_j$, any positive 
instance $x$ must differ from $y$ on position 1 or position $j$ (or both).  
Thus $\max_{i, x_i \ne y_i} p_x(i)p_y(i) \ge 1/2n$ and 
$\mpi(\Col) \le \sqrt{2n}$.

Now for the lower bound.
Fix a set of probability distributions $p_x$.  Let $x$ be any positive
instance.  There must be at least $n/2$ positions $i$ satisfying
$p_x(i) \le 2/n$.  Call this set of positions $I$.  Now consider a negative
instance $y$ of where $y_j=x_j$ for all $j \not \in I$, and $y$ is assigned
values in $I$ in an arbitrary way so as to make it a negative instance.
For this pair $x,y$ we have $\max_i \sqrt{p_x(i)}\sqrt{p_y(i)} \le \sqrt{2/n}$,
thus $\mpi(\Col) \ge \sqrt{n/2}$.
\end{proof}
                                                                                
The following table summarizes the bounds from this section.
%
%
\setlength{\tabcolsep}{2pt}
\begin{center}
\begin{tabular}{|l|c|c|c|c|c|c|}
\hline
Function&Input & $\mathsf{sum}$ &$Q_\epsilon$ & $\mathsf{max}$ & $\L$ & $s_0s_1$ \\ 
& size&$\mathsf{PI}$&&$\mathsf{PI}$&&\\
\hline\hline
$\scriptstyle {\MAJ{h}}$      & $N$  & $2^h \approx$ & $\scriptstyle \Omega(N^{0.63})$ & $\scriptstyle N^{0.63}$ & $\scriptstyle \Omega(N^{1.26}), $ & $\scriptstyle N^{1.26}$\\

& $=3^h$& $N^{0.63}$ &                    &            & $\scriptstyle O(N^{1.46})$         &
\\ \hline
$\A^h$      &  $N$ & $\scriptstyle 2.5^h\approx$            & $\scriptstyle \Omega(N^{0.66})$& $\scriptstyle \leq 3^h\approx$         & $\scriptstyle \Omega(N^{1.32}), $ & $N$ \\
            &   $=4^h$     & $ \scriptstyle N^{0.66}$ & {\small \cite{A03} }       & $\scriptstyle N^{0.79}$ &  $\scriptstyle O(N^{1.79}) $& 
\\ \hline
$\Col$      &  $N$  & 2 & $\scriptstyle \Theta(N^{1/3})$ & $\scriptstyle \Theta(\sqrt{N})$  & $N$ & $\bot$
\\ \hline 
\end{tabular}
\end{center}

\section{Conclusions and open problems}

Our new formula size lower bound
techniques subsume many previous techniques, and for
some functions they are provably better.
A significant part of our intuition comes from
quantum query complexity and Kolmogorov complexity. Measures $\spi$ and
$\mpi$ have many interesting properties and they connect different complexities
 such as quantum query complexity, classical formula size, 
classical probabilistic formula size and circuit depth. 

An outstanding open problem is whether the square of the quantum query
complexity lower bounds the formula size. 
Another is that we do not know a nice dual
expression for $\mpi$, and it does not seem to be a natural property
in the sense of Razborov and Rudich. Thus the study of $\mpi$ may 
lead us to a better understanding of complexity measures that
themselves are hard to compute. We could reprove a key lemma
of H{\aa}stad that leads to the best current formula size lower bound and we
are hopeful that our techniques eventually will lead to improvements of the
bounds in \cite{H98}.

\section*{Acknowledgments}
We would like to thank Fr\'ed\'eric Magniez, Robert \v{S}palek, and Ronald
de Wolf for helpful discussions.  We also wish to thank Ryan O'Donnell for
suggesting to look at the recursive majority of three function, and 
Xiaomin Chen for help in programming.  Finally, we thank the anonymous referees 
for many exposition improving comments.


\newcommand{\etalchar}[1]{$^{#1}$}

\end{document}